\documentclass[twocolumn,prb,10pt,aps,longbibliography,superscriptaddress]{revtex4-2}
\usepackage{graphicx}
\usepackage{amsmath}
\usepackage{amssymb}
\usepackage{amsfonts}
\usepackage{dsfont}
\usepackage{dcolumn}
\usepackage{bbm}
\usepackage{bm}
\usepackage{tikz}
\usetikzlibrary{positioning}
\usepackage{mathtools}
\usepackage{braket}
\usepackage{physics}
\usepackage{siunitx}
\usepackage[unicode]{hyperref}
\usepackage[nameinlink]{cleveref}

\usepackage{bbold} 
\usepackage{nicefrac}
\usepackage{tikz}
\usetikzlibrary{arrows.meta}
\usepackage{xcolor}

\usepackage[makeroom]{cancel}

\newcommand{\bes}{\begin{subequations}}
\newcommand{\ees}{\end{subequations}}
\newcommand{\be}{\begin{equation}}
\newcommand{\ee}{\end{equation}}

\newcommand{\uu}[1]{\underline{\underline{#1}}}
\newcommand{\sw}{^\text{sw}}
\newcommand{\thr}{^\text{thr}}

\newcommand{\sumnn}{\sum_{\langle i,j\rangle}}

\newcommand{\lr}[1]{\left( #1 \right)}

\newcommand{\lara}[1]{\langle #1 \rangle}

\newcommand{\nbi}[1]{#1_i^\dagger {#1}_i}
\newcommand{\nbk}[1]{#1_\mathbf{k}^\dagger {#1}_\mathbf{k}}

\newcommand{\sumk}{\sum_{\mathbf{k}}}

\newcommand{\ak}{a_\mathbf{k}}
\newcommand{\bk}{b_\mathbf{k}}
\newcommand{\akm}{a_\mathbf{-k}}
\newcommand{\bkm}{b_\mathbf{-k}}

\newcommand{\gam}{\gamma_\mathbf{k}}

\begin{document}
    \title{Exchange enhanced switching by alternating fields in quantum antiferromagnets}


    \author{Asliddin Khudoyberdiev}
    \email{asliddin.khudoyberdiev@tu-dortmund.de}
    \affiliation{Condensed Matter Theory, 
    TU Dortmund University, Otto-Hahn-Stra\ss{}e 4, 44221 Dortmund, Germany}

    \author{G\"otz S.\ Uhrig}
    \email{goetz.uhrig@tu-dortmund.de}
    \affiliation{Condensed Matter Theory, 
    TU Dortmund University, Otto-Hahn-Stra\ss{}e 4, 44221 Dortmund, Germany}

    \date{\textrm{\today}}

 \begin{abstract}
Information can be stored magnetically in antiferromagnets ultrafast since their 
characteristic times are on the picosecond  scale. Various spin 
torques have proven to be important for efficient and high-speed magnetic memories. 
So far, this has been understood on the classical level by solving the equations
of motion for macrospins describing the collective motion of the sublattice
magnetizations. Since spins and hence magnetizations are deeply rooted 
in quantum mechanics, we show that the exchange enhanced manipulation of
sublattice magnetizations extends to quantum antiferromagnets as well. 
To this end, we solve the time-dependent mean-field equations for Schwinger boson
theory under external alternating magnetic fields. Exchange enhancement
persists on the quantum level which includes dephasing effects.
Significantly lower fields are sufficient to control the sublattice magnetization 
than for uniform fields which holds great promises for the realization of 
ultrafast magnetic storage devices.
\end{abstract}

    \maketitle

		
One requirement for ultrafast storage devices is that they can be operated in the terahertz (THz) regime and that they provide a large storage capacity. Advancements in understanding and manipulating antiferromagnetic order paves the way for innovative technologies with 
disruptively improved performance.  
Antiferromagnets are promising candidates to speed up information processing because of their 
characteristic frequencies range in the THz regime \cite{jungw16}. Additionally, antiferromagnetic domains exhibit hardly any stray fields because their net magnetization cancels; this enables one 
to reduce the distance between domains encoding bits \cite{loth12}. Nevertheless, the efficient  control of antiferromagnetic order remains one of the main challenges for applications. In order to have bits robust against perturbations
anisotropic spin systems are considered. In return, their manipulation
requires to overcome  activation energies \cite{bolsm23,khudo24a} for switching between the 
favored spin states \cite{gomon17,song18}.

\begin{figure}[htb]
    \centering
    \includegraphics[width=\columnwidth]{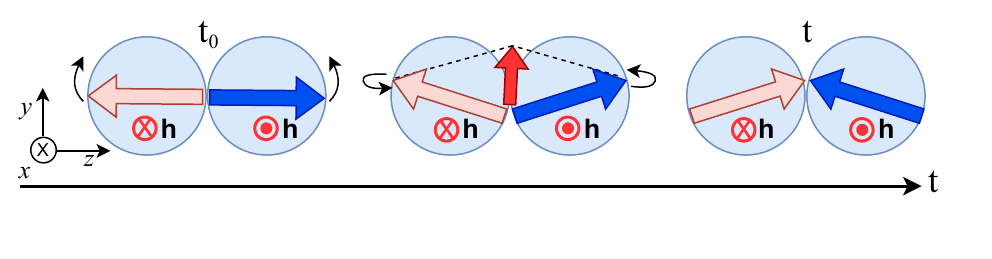}
   \caption{Sketch of exchange enhanced switching by an alternating field. 
	Light red and blue arrows stand for the magnetizations on sublattices A and B.
	The alternating external field ${\bf h}$ tilts the magnetization in opposite 
	directions so that a net magnetization is induced (dark red arrow) 
	and thereby an internal exchange field around which	both sublattice magnetizations
	quickly precess so that the N\'eel vector is essentially rotated by $180^\circ$. This illustration
	is a simplification because tilting and precessing happen simultaneously.}
    \label{fig:ee}
\end{figure}

In the first place, efficient switching means that only low external fields need
to be employed. In experimental and in classical macrospin descriptions it turned out
\cite{gomon10,wadle16,roy16} that it is advantageous to exploit so-called exchange enhancement.
Since the internal fields exerted by adjacent spins via exchange coupling are by far larger
than the external ones it is beneficial to have them assist in the reorientation
of the N\'eel vector. This is achieved
by slightly canting the antiparallel sublattice magnetizations so that a net
magnetization is induced. Then, the sublattice magnetizations are precessing 
around the internal magnetic field stemming from the net magnetization, see  Fig.\ \ref{fig:ee}.

At first sight, the required alternating magnetic field may appear elusive, but
theoretical \cite{zelez14} and experimental progress \cite{olejn18,bodna18} in the last decade has shown that current-induced spin-orbit torques can be of N\'eel type, i.e., they are alternating 
between the two sublattices. Recently,  Behovits {\it et al.} \cite{behov23} even achieved
a deflection of the N\'eel vector in Mn$_2$Au by up to 30$^{\circ}$. Besides current induced
torques one can also conceive bipartite systems in which the $\uu{g}$ tensor is anisotropic due 
to large spin-orbit couplings so that it is different $\uu{g}|_A \neq \uu{g}|_B$
between both sublattices. Then, even a uniform magnetic field $\vec B$ of a THz pulse generates an alternating field  $\vec h_a = (\uu{g}|_A - \uu{g}|_B)\mu_{\text B}\vec B/2$.

It is the present key objective to investigate the exchange enhanced
control on the quantum level. We confirm that the characteristic energy 
is not $h_a$, but $\sqrt{J h_a}$ where $J$ is the exchange coupling 
\cite{kitte51,kimel04,gomon16,gomon18}. This goal
is achieved by using a time-dependent Schwinger mean-field theory which we have developed
very recently \cite{bolsm23,khudo24a}. So far, it has been applied successfully to describe the
effects of {\it uniform} static \cite{bolsm23} and time-dependent external fields \cite{khudo24a}.
Its two main assets compared to classical macrospins 
are (i) to capture leading quantum fluctuations and (b) dephasing
since all spin modes contribute at their respective frequencies \cite{khudo24a}.
The spin gap due to the spin anisotropy determines the required 
threshold field strength necessary to overcome the anisotropy potential barrier in
reversing the orientation of the sublattice magnetization. The switching time, i.e., the 
time required to achieve the reorientation is inversely proportional to the external uniform
field, i.e.,  $t\sw \propto 1/h$.
Here, we extend this analysis  to the relevant case of alternating fields. The particularly promising finding is that much smaller fields are sufficient for switching
and that the switching dynamics is still ultrafast.

Concretely, we consider the easy-axis spin-1/2 Heisenberg model at zero temperature on 
a simple cubic lattice\footnote{Results for the square lattice
 are included in the Supplemental Material \cite{suppl}.} 
with nearest-neighbor interactions
\be
  \mathcal{H}_0 = J \sumnn \big\{\frac{\chi}{2}
	\lr{S^+_iS^-_j + S^-_iS^+_j} + S^z_i S^z_j \big\}  ,
	\label{eqn:aniAFM-H2}
\ee
where $J_x=J_y=J_{xy}$, $J_z=J$ and $\chi=J_{xy}/J_z$.
The Zeeman term for an alternating time-dependent magnetic field reads
\begin{equation}\label{eq:alternatH}
    \mathcal{H}_{\text{alt}} = - \vec{h}_\text{a}(t)\cdot\sum_i (-1)^i \vec{S}_i.
\end{equation}
For technical reason, we specify $\vec{h}_\text{a}$ to point along the $x$ axis.
We use $J$ as energy unit.

Standard spin wave theory according to Holstein-Primakoff or Dyson-Maleev
only expands in fluctuations around the ordered state which is not appropriate for
capturing reorientations by $180^\circ$. Thus, we employ the the Schwinger boson representation
\be
\label{eqn:SB-AFM}
        S_i^+ = a_i^\dagger b_i , \quad 
        S_i^- = b_i^\dagger a_i , \quad       
S_i^z  = \frac{1}{2}\big(\nbi{a} - \nbi{b}\big)
\ee
with two bosons per site and the constraint $2S=\nbi{a} + \nbi{b}$ which is fulfilled on average
in the mean-field approach \cite{auerb88,auerb94}. The sublattice magnetization
in $z$ direction reads 
\begin{equation}
\label{eq:magnetization}
   m = \frac{1}{2}\big(\langle \nbi{a} \rangle  - \langle \nbi{b}\rangle\big) .
\end{equation}
All equations simplify by a sublattice rotation as it is common for
antiferromagnets. We rotate all spins on the B lattice by 180$^\circ$ about $S_i^y$
which implies for the Schwinger bosons $a_j \rightarrow -b_j$ and $b_j \rightarrow a_j$.
Since this results $S^x\rightarrow -S^x$ on sublattice B, the alternating external
field in $x$ direction becomes uniform \footnote{A uniform control field is realized
by applying the field in $y$ direction \cite{bolsm23,khudo24a}.} 
so that the Hamiltonian after the sublattice rotation is translationally invariant and reads
\be
  \mathcal{H} = J \sumnn \big\{\frac{\chi}{2}
	\lr{S^+_iS^+_j + S^-_iS^-_j} + S^z_i S^z_j \big\}  
	-h_{\mathrm{a}}(t)\sum_i S^x_i.
	\label{eqn:aniAFM-H3}
\ee
This is re-expressed in Schwinger bosons according to \eqref{eqn:SB-AFM}. The mean-field
Hamiltonian is obtained by introducing the complex expectation values
$A\coloneqq \lara{a_i a_j + b_i b_j}$ and $B\coloneqq \lara{a_i a_j - b_i b_j}$ and
applying Wick's theorem.
After Fourier transformation the bilinear mean-field Hamiltonian reads 
\begin{align}
\nonumber
   \mathcal{H}_\text{MF} &= E_0 -\frac{z}{8}\sumk \gam \big(C_-\ak^\dagger \akm^\dagger 
	+ C_+\bk^\dagger\bkm^\dagger 
	\\ \nonumber
	& \quad + C_-^*\ak\akm + C_+^*\bk\bkm\big) + \lambda \sumk \big(\nbk{a} + \nbk{b}\big) 
	\\
	& \quad - \frac{1}{2} h_{\mathrm{a}}(t)\sumk\big(\ak^\dagger\bk + \bk^\dagger\ak\big) ,
	\label{eq:hamilton-switch}
\end{align} 
where $z$ is the coordination number and $C_{\pm}\coloneqq A(1+\chi)\mp B(1-\chi)$. The energy 
$E_0$  does not contribute to the dynamics of the system so that we omit it henceforth. The wave vector only enters via 
\be
\gamma_k=\frac{1}{d}\sum_{i=1}^d\cos{k_i},
\ee
where $d$ is dimension of the lattice and the lattice constant is set to unity.

The initial conditions are found by Bogoliubov transformation of the boson and self-consistently
determination of  $A$,  $B$, and other expectation values, 
see Refs.\ \cite{auerb88,auerb94,bolsm23}. We follow this route
as well, but finally transform the results back to the  expectation
values  $\lara{\ak \akm}$, $\lara{\bk \bkm}$ and of their conjugates as
well as of $\lara{\nbk{a}}$ and $\lara{\nbk{b}}$, see Supplement \cite{suppl}. Since only
 $\gamma_k$ matters, it is sufficient to discretize the interval $\gamma\in[-1,1]$
and to determine for the discrete values $\lara{\ak \akm}_\gamma$, $\lara{\bk \bkm}_\gamma$,
$\lara{\nbk{a}}_\gamma$ and $\lara{\nbk{b}}_\gamma$. This simplifies the numerics greatly and makes
high-precision computations in three dimension possible; the required densities-of-states
are given in the Supplement \cite{suppl} where the Heisenberg equations of motions resulting 
from the mean-field Hamiltonian $\mathcal{H}_\text{MF}$ are also provided.

Control is achieved by an external magnetic field. This can
be done by a static, constant field $h_\text{static}$ or by a pulse of finite duration and maximum
amplitude $h_\text{pulse}$. We study both variants, but expect from previous
results \cite{bolsm23,khudo24a} that pulses are
 more efficient if the pulse frequency is close to resonance
with the spin gap. THz pulses are advantageous anyway in view of experimental feasibility
 \cite{kampf11}. We consider the alternating pulse (subscript `a')
\begin{equation}
\label{shortpulse}
h_\mathrm{a}(t)=
h_\mathrm{a,pulse}\cos(\alpha\Delta (t-3\tau)+\phi_0)\cdot e^{-\frac{(t-3\tau)^2}{2\tau^2}} 
\end{equation}
where $h_\text{a,pulse}$ is the maximum amplitude, $\Delta$ is the spin gap, $\alpha < 1$ is a  renormalization factor to optimize the resonance, and $\phi_0$ is a phase shift. 
The pulse duration is given by $\tau$. Since we start the simulations at $t=0$ we shift
the pulse by $3\tau$ to capture it fully. The analysis and optimization of the pulse parameters in \eqref{shortpulse} are given in Supplemental Material \cite{suppl}. It turns out that the same parameters as for uniform fields 
\cite{khudo24a} yield good results. Hence we use again 
$\alpha=0.85$, $\tau=10 \,J^{-1}$, and $\phi_0=\pi/3$.

\begin{figure}[htb]
    \centering
    \includegraphics[width=\columnwidth]{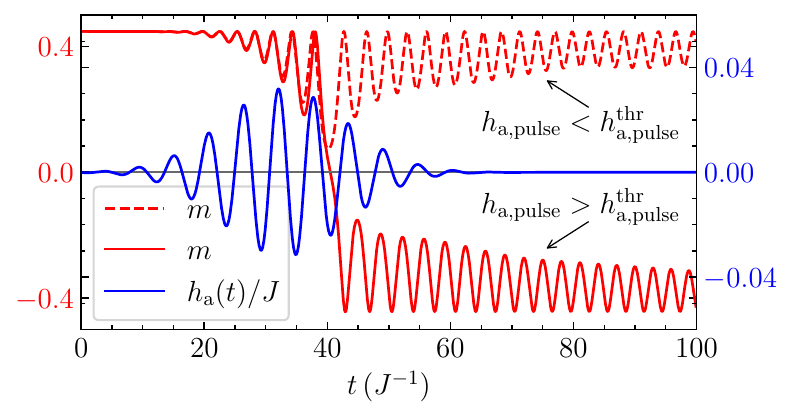}
   \caption{Evolution of the sublattice magnetization $m(t)$ for  
	$h_\mathrm{a,pulse}=0.031 \, J$ (dashed red line) and $h_\mathrm{a,pulse}=0.032 \,J$ 
	(solid red line) at $\chi=0.9$ where $h\thr_\text{a,pulse}=0.0313 \,J$. 
	The values of the pulse with $h_\mathrm{a,pulse}=0.032 \,J$ (blue line) 
	are denoted on the right $y$ axis. }
    \label{fig:pulsemag}
\end{figure}

Figure \ref{fig:pulsemag} displays two representative behaviors of the sublattice magnetization
subject to pulses of the type \eqref{shortpulse}. There is a threshold value of the amplitude
which has to be overcome to realize switching that is clearly signaled by the sign change
of $m(t)$. If the pulse has too low amplitude only oscillations close to the initial
magnetization are induced. A large enough amplitude nudges the magnetization over its 
anisotropic maximum so that it oscillates thereafter close to  its negative equilibrium value.
The oscillations as such are not surprising since the pulse perturbs the system injecting energy
so that oscillations are induced. We emphasize that no relaxation is included here
 since we  want to study the closed quantum system.
In view of the absence of relaxation one may wonder why the oscillations decrease at all.
Indeed, typical classical macrospin calculations without relaxation display
persisting sign changes; no decrease in oscillations is observed \cite{khudo24a}.
The observed decrease in the quantum model results from dephasing. Many modes contribute
to collective observables such as $m(t)$ with their individual frequencies so that
the increasing phase differences lead to a decreasing total signal. This is
clearly visible in Fig.\ \ref{fig:pulsemag} and was to be expected from previous
quantum calculations \cite{bolsm23,khudo24a}. Relaxation will surely speed up the 
decrease of the oscillations. But its  quantitative investigation 
is beyond the scope of the present study.
Here, we conclude  that the applied amplitude of the pulse is sufficiently low to reach 
the realizable range, e.g.,  $h_\mathrm{0a}=0.032 \, J$ correspond to about 2.8 T for antiferromagnetic coupling constant $J=10 \, \si{\meV}$.  
The threshold value decreases further for weaker anisotropy $\chi\to1$.

\begin{figure}[htb]
    \centering
    \includegraphics[width=\columnwidth]{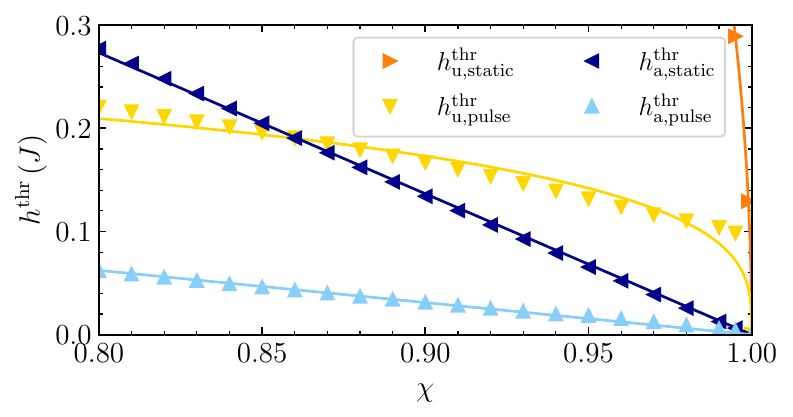}
   \caption{Threshold fields vs.\ anisotropy $\chi$ for the simple cubic lattice.
	The fits (solid lines) for uniform fields use 
	$h\mathrm{_{u}^{thr}}=c_u(1-\chi^2)^n$  where for static case 
	we find $c_u\approx 2.96 \, J$, $n\approx 0.504\pm 0.585$ and for the pulse  
	$c_{u}=\approx 0.28 \, J$, $n=0.30\pm 0.02$. For alternating fields, 
	the behavior is clearly linear which we fit by
 $h\mathrm{_{a}^{thr}}=c_{a}(1-\chi)$, where $c_{a}\approx 1.365\, J$ for the static 
and $c_{a}\approx 0.312 \, J$ for the time-dependent case.}
    \label{fig:threshchi}
\end{figure}

Figure \ref{fig:threshchi} depicts our key result. The threshold values labeled
by the superscript $\thr$ for the four variants
considered are plotted. The corresponding amplitudes are $h_\text{u,static}$ for
 a uniform, constant field, $h_\text{u,pulse}$ for a uniform pulse, $h_\text{a,static}$ for an alternating, constant field, and  $h_\text{a,pulse}$ for an alternating pulse. The crucial
observation is the very different power law behavior for $\chi\to1$. For uniform, static
field we had established that the threshold is almost quantitatively given by the spin gap
implying $h\thr \propto \sqrt{1-\chi}$, see orange line in Fig.\ \ref{fig:threshchi}. 
Such a sublinear power law is also indicated
by the results given by the yellow line. The uniform static case
requires by far the largest fields. The alternating static and pulse case display linear
behavior which implies the highly advantageous feature 
that their thresholds become much smaller than the uniform ones for weak anisotropies 
$\chi\to 1$. This is a direct consequence of the exchange enhancement. The activation energy to overcome is given by the spin gap $\Delta \propto \sqrt{1-\chi}$.
If the relevant energy scale of the control field is $\sqrt{J h_\text{a}}$ instead of $h_\text{a}$
the ensuing threshold is linear in $1-\chi$ 
\be
\Delta \propto J\sqrt{1-\chi} \propto \sqrt{J h\thr_\text{a}} 
\quad \Rightarrow\quad  h\thr_\text{a} \propto J(1-\chi).
\ee
For an estimate, the threshold at $\chi=0.995$ is 
$h\thr_\text{a,pulse}=0.0041 \, J$ corresponding to about 
$0.5\,$T only at $J\approx 14 \,\si{\meV}$.  These observations
in a quantum model show strikingly the advantages of using alternating pulses for
the re-orientization of magnetizations.

\begin{figure}[htb]
    \centering
    \includegraphics[width=\columnwidth]{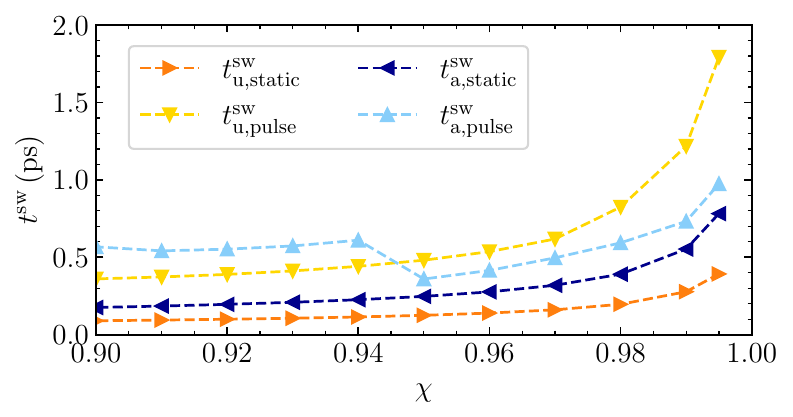}
   \caption{Switching time vs.\  the anisotropy for fields 20\% above the values shown in
	Fig.\ \ref{fig:threshchi}. For the pulses, we measure $t\sw$ not from $t=0$, but from the center of the pulse, i.e., we deduct $3\tau$.}
    \label{fig:tswitchchi}
\end{figure}

The next key quantity to study is the switching time, i.e., the time required for the manipulation. Low switching fields which in turn lead to very long switching times $t\sw$
would not be of great help in view of applications. Thus, we analyze the time 
it takes for the relevant sign change in $m(t)$ to occur if we switch the system
with an amplitude 20\% above the threshold. Using this increased values 
is numerically more stable than a a study at the marginal field amplitude.
In any application one would surely use  sufficiently large fields for the control. 
Figure \ref{fig:tswitchchi} shows the switching time as a function of the anisotropy. 
For concreteness, we provide times in picoseconds assuming an exchange coupling of 
$J\approx10 \si{\meV}$. At first glance, Fig.\ \ref{fig:tswitchchi} 
does not convey a clear message because the pulses take longer than the static fields
and the quickest switch is not obtained by the alternating field, but by the uniform one.
But one has to keep in mind that the employed control fields are very different, namely
 much smaller for the pulses than for the static fields. Hence, the message
from Fig.\ \ref{fig:tswitchchi} is not the difference between the different variants,
but their similarity: despite the largely different fields the switching times 
are in the range of $\approx 1\,$ps. This confirms that the THz range is the
appropriate range of magnetization re-orientation in quantum antiferromagnets.

\begin{figure}[htb]
    \centering
    \includegraphics[width=\columnwidth]{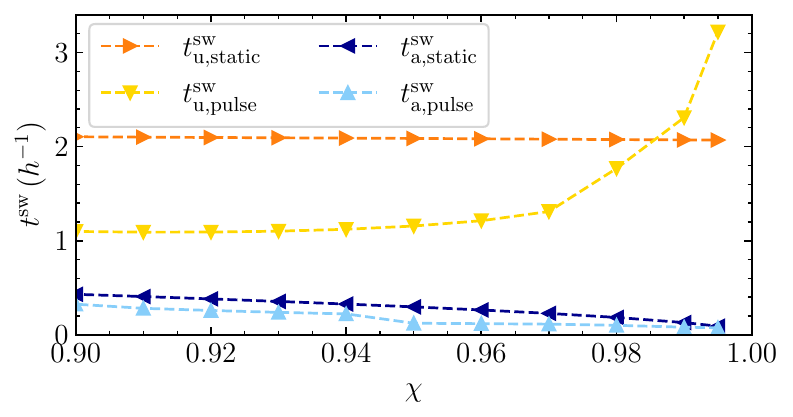}
   \caption{Switching times of the four variants expressed relative to the applied
	field amplitude $h=1.2h\thr$.}
    \label{fig:tswitchcompare3D}
\end{figure}

To underline that the switching speed must be
seen relative to the employed field amplitudes we plot the data from 
Fig.\ \ref{fig:tswitchchi} in Fig.\ \ref{fig:tswitchcompare3D} in units of
$1/h$ \cite{bolsm23}. This rendering clearly shows that the alternating fields
are more efficient than the uniform ones. Also, the pulses
are  advantageous relative to constant fields.
For the uniform, static case an almost constant result is observed
implying that for sufficiently large fields only the field amplitude
determines the dynamics. For the uniform pulse, 
we find indications of a weak divergence of the switching time showing
that the some power of the inverse spin gap determines $t\sw$.
This is qualitatively consistent with the weak decrease of $h\thr_\text{u,pulse}$
we observed in Fig.\ \ref{fig:tswitchchi}.
The two blue lines corresponding to the alternating fields show decreasing
values of $h t\sw$ for $\chi$ tending to 1 which reflects a proportionality
$\propto \sqrt{1-\chi}$. This is understood from the estimate
\be
h\thr_\text{a} t\sw_\text{a} \propto \frac{J (1-\chi)}{\sqrt{J h\thr}} \propto 
\frac{J (1-\chi)}{\Delta} \propto \sqrt{1-\chi}.
\ee
Hence, it is not surprising that the alternating control field provides the
fastest control in spite of using the smallest control fields.
This is a very promising observation in view of experimental
realizations.
	
On the way to using magnetic degrees of freedom for data storage and handling
antiferromagnetism is a realm to be fully understood, in particular the possibilities
of control by external fields. This is required for writing and erasing data.
We investigated the effect of alternating control fields on the orientation of the 
sublattice magnetization in an easy-axis Heisenberg quantum antiferromagnet on a simple
cubic lattice with numerical results for $S=1/2$. The fields alternate in orientation
between the two sublattice and they are assumed to be static or in pulse shape with
a carrier frequency of slightly below resonance to the spin gap. The employed tool
was time-dependent Schwinger boson mean-field theory because it is the only mean-field 
approach able to describe full revolutions in time of the magnetic order as well as the
static equilibrium while  maintaining the assets of a quantum model.

In  line with previous classical and experimental evidence, our results clearly show 
that alternating effective magnetic fields, also called N\'eel spin torques, are 
the best choice for a rapid manipulation of the expectation value of the sublattice
magnetization. Much lower fields are needed than for uniform fields because
of the exchange enhancement which implies that the characteristic energy of the
control field is not $h$, but $\sqrt{J h}$ where $h=g\mu_\text{B}B$.
Very importantly, also the time required for performing the manipulation does not
grow too much. Even for very weak fields we predict a switching time in the 
picosecond range. The required fields can be as low as tenths of Tesla.
For $\chi=0.999$, we estimate $h\thr_\text{a,pulse}=0.0021 \,J$; this corresponds roughly
to $\approx 0.2\,$Tesla assuming $J\approx 10 \si{\meV}$ for the simple cubic lattice. In summary, these results
and the method developed to obtain them pave the way to a better 
understanding of magnetization dynamics and hence a sustainable information processing
based on quantum antiferromagnetism.

Next steps of a theoretical analysis suggesting themselves are the inclusion
of relaxation on the quantum level \cite{uhrig24}. The spin size $S$ can very
easily be enhanced which in turn allows to consider more complex anisotropies,
for instance with four-fold equilibrium orientations of the magnetization.
Finally, other bipartite lattice can be addressed as well in a quite straightforward 
manner. Hence, there is a plethora of systems and issues to be studied in the
near future.

		

We are grateful for helpful discussion with Tobias Kampfrath.
This work has been financially supported by the Deutsche
Forschungsgemeinschaft (German Research Foundation)
in project UH 90/14-1 and by the Stiftung Mercator in
project Ko-2021-0027.
    



%

\section{Supplemental Material}
\label{app:den}
\subsection{ Calculation of density-of-states of $\gamma_k$}

To consider long range order in the system, it is important to choose the lattice size large enough. However, possible points in Brillouin zone becomes large especially for 3D simple cubic lattice and the sums in the Hamiltonian in momentum space will be numerically unmanageable.  Hence, instead of summing over each value of momentum in first Brillouin zone, one can convert the sums into one dimensional integrals as
\be
\lim_{N\rightarrow \infty}\frac{1}{N}\sum_kF(\gamma_k)=\int_{-1}^1d\gamma\rho(\gamma)F(\gamma)
\label{sup:sumtoint}
\ee
where $\rho(\gamma)$ is density of state in $d$ dimension. It has the following forms for square and simple cubic lattices \cite{hanis97}
\bes
\label{sup:density2D}
\begin{align}
&\rho_{\mathrm{sq}}(\gamma)=\frac{2}{\pi^2}K(1-\gamma^2), \label{sup:density3D}\\
&\rho_{\mathrm{cub}}(\gamma)=\frac{1}{\pi}\int_{u_1}^{u_2}\frac{du}{\sqrt{1-u^2}}\rho_{\mathrm{sq}}(\gamma+u/3)\label{sup:densities}, \\
& u_1=\max(-1,-2-3\gamma),  \quad  u_2=\min(1,2-3\gamma).
\end{align}
\ees
We calculated these integrals with Newton-Cotes midpoint interval rule. Now, to calculate the integral in \eqref{sup:sumtoint}, one can discretize it for possible $\gamma$ points and solve differential equations, obtained from Heisenberg's equation motion in the main text, for each values of $\gamma$ where $\gamma\in[-1,1]$. We have chosen 2000 gamma points in the given interval and the results are consistent with sum method for sufficiently large lattice size. 


\subsection{Equations for the non-equilibrium dynamics of expectation values}
\label{app:Eqs}
The mean-occupation numbers of $a$ and $b$ bosons can be calculated in the equilibrium within the process of Hamiltonian diagonalization 
\bes
\label{eqn:exp-values}
\begin{align}
\lara{\nbk{a}}_{\gamma} &= \frac{\lambda}{2\omega_k^-(\gamma)}-\frac{1}{2}   , \label{nakg} 
\\
\lara{\nbk{b}}_{\gamma} &=  \frac{\lambda}{2\omega_k^+(\gamma)}-\frac{1}{2} \label{nbkg}.
\end{align}
\ees

Here, the spin wave dispersion relations for $\alpha$ and $\beta$ bosons $\omega_k^-$ and $\omega_k^+$, respectively, reads 
\be
\label{dispersion}
    \omega^{\pm}_\mathbf{k} = \sqrt{\lambda^2-\lr{{z|C_\pm|\gam}/{4}}^2}.
\ee
Due to the anisotropy, both boson dispersions gain a energy gap as 
\be \label{eq:energygap}
\Delta^\pm \coloneqq  \omega^\pm_{\textbf{k}=0}. \quad  \Delta=\Delta^+-\Delta^- ,
\ee
where $\Delta$ is the physical spin gap. 

 The variables $A$ and $B$ required to compute $C_\pm$ in \eqref{eq:hamilton-switch} are defined by
\bes
\label{eq:ABS-compute}
\begin{align}
 A &= \lara{a_ia_j} + \lara{b_ib_j}
\\
 &=\int_{-1}^1 \gamma\rho_d(\gamma)\lr{\lara{\ak\akm}_{\gamma}
+\lara{\bk\bkm}_{\gamma}}d\gamma , \label{eqn:IV3} 
\\ 
 B &= \lara{a_ia_j} - \lara{b_ib_j} 
\\
&=\int_{-1}^1 \gamma\rho_d (\gamma)\lr{\lara{\ak\akm}_{\gamma}-\lara{\bk\bkm}_{\gamma}}d\gamma  ,\label{eqn:IV4}
\\
2S &= \lara{\nbi{a}} + \lara{\nbi{b}}
\\
&= \int_{-1}^1\rho_d(\gamma) \big(\lara{\nbk{a}}_{\gamma} 
+ \lara{\nbk{b}}_{\gamma}\big)d\gamma  .\label{eqn:IV5}
\end{align}
\ees
The last equation responsible to fulfill the constraints on boson number and 
the density $\rho_d(\gamma)$ is the density-of-states in $d$ dimensions 
for $\gam$  \cite{hanis97}. The other expectation values are
\bes
\label{eqn:aabb}
\begin{align}
\lara{\ak\akm}_{\gamma} &= \frac{z\gamma C_-}{8\omega_k^-(\gamma)} ,\label{bbkg} 
\\ 
\lara{\bk\bkm}_{\gamma} &= \frac{z\gamma C_+}{8\omega_k^+(\gamma)}  .\label{aakg}
\end{align}
\ees

Finally, the temporal evolution is determined from the equations
of motion for the introduced expectation values. 
As stated before, this dynamics only depends on the value $\gam=\gamma$
\bes
\label{eqn:DissEQ}
\begin{align}
    \partial_t \lara{\ak^\dagger \ak}_{\gamma} &= -i\frac{z}{4} 
		\gamma \big(C_-^*\lara{\ak\akm}_{\gamma}-
		C_-\lara{\ak^\dagger\akm^\dagger}_{\gamma}\big) 
		\nonumber \\
	 & \quad + i\frac{h_x}{2}\big(\lara{\ak^\dagger\bk}_{\gamma}-
		\lara{\bk^\dagger\ak}_{\gamma}\big), \label{eqn:DissEQ1}
\\
    \partial_t \lara{\bk^\dagger \bk}_{\gamma} &=-i \frac{z}{4} 
		\gamma \big(C_+^*\lara{\bk\bkm}_{\gamma}
		-C_+\lara{\bk^\dagger\bkm^\dagger}_{\gamma}\big) 
		\nonumber \\ 
    &\quad - i\frac{h_x}{2}\big(\lara{\ak^\dagger\bk}_{\gamma}-
		\lara{\bk^\dagger\ak}_{\gamma}\big), \label{eqn:DissEQ2}
\\
    \partial_t \lara{\ak \akm}_{\gamma} &=  i \frac{z}{4}\gamma 
		\big[C_- (2\lara{\nbk{a}}_{\gamma}+1)\big]
		\nonumber \\
		&\quad - 2 \lambda  i \lara{\ak\akm}_{\gamma} 
		+ ih_x\lara{\ak\bkm}_{\gamma} ,\label{eqn:DissEQ4}
\\
    \partial_t \lara{\bk \bkm}_{\gamma} &=  i\frac{z}{4} \gamma 
		\big[C_+ (2\lara{\nbk{b}}_{\gamma}+1)\big]
		\nonumber \\
		&\quad - 2 \lambda  i \lara{\bk\bkm}_{\gamma} 
		+ih_x\lara{\ak\bkm}_{\gamma} ,\label{eqn:DissEQ5}
\\
    \partial_t \lara{\ak^\dagger \bk}_{\gamma} &= - i \frac{z}{4}\gamma 
		\big(C_-^*\lara{\ak\bkm}_{\gamma} 
		- C_+\lara{\ak^\dagger\bkm^\dagger}_{\gamma}\big)
		\nonumber \\
		&\quad-i \frac{h_x}{2}\big(\lara{\bk^\dagger\bk}_{\gamma}
		-\lara{\ak^\dagger\ak}_{\gamma}\big) ,\label{eqn:DissEQ3}
\\
    \partial_t \lara{\ak \bkm}_{\gamma} &=  i \frac{z}{4} \gamma 
		\big( C_-\lara{\ak^\dagger\bk}_{\gamma} + 
		C_+ \lara{\bk^\dagger\ak}_{\gamma} \big)
		\nonumber \\ \nonumber
     &\quad -2 \lambda  i \lara{\ak\bkm}_{\gamma}\\
		 &\quad + i\frac{h_x}{2}\big(\lara{\ak\akm}_{\gamma}+\lara{\bk\bkm}_{\gamma}\big).
				\label{eqn:DissEQ6}
\end{align}
\ees
These above set of differential equations are solved with calculated initial values from \eqref{eqn:exp-values} and \eqref{eqn:aabb}.


\subsection{The analyses of the pulse for a simple cubic lattice}
\label{app:3Dtime}
\begin{figure}[htb]
    \centering
    \includegraphics[width=\columnwidth]{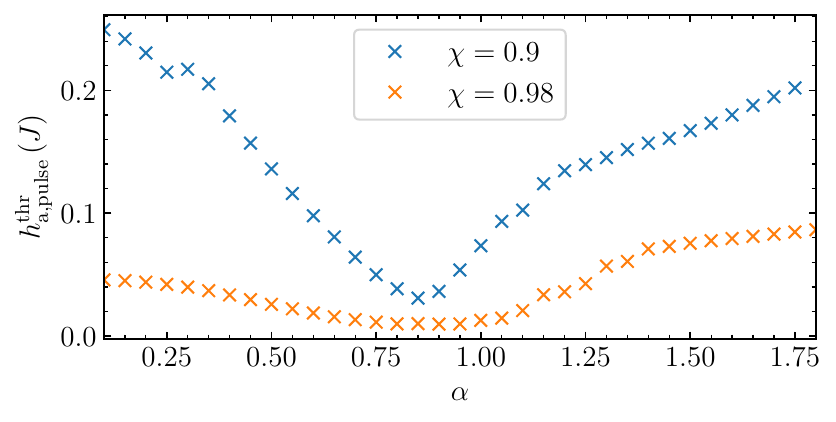}
   \caption{Threshold field dependence on frequency renormalization constant in \eqref{shortpulse}. The other parameters are $\phi_0=\pi/3$ and $\tau=10 \, J^{-1}$.}
    \label{fig:threshalpha}
\end{figure}

Figure \ref{fig:threshalpha} shows the essence of the resonance where $h\mathrm{^{thr}_{a,static}}$ is the threshold amplitude of the pulse in Eq. \eqref{shortpulse} to switch antiferromagnetic order. The perturbation energy should be in the order of spin gap $\Delta$. In particular, slight deviation from the frequency $\omega=\Delta$ occurs and the anisotropy gap decreases around switching process due to different level separations between the states \cite{miyas23}. Hence, frequency renormalization constant $\alpha$ is helpful to catch the best resonance coupling. According to our test results,  $\alpha\approx 0.85$  is the optimum value for valid $\chi$ with lowest threshold field as in Ref.\ \cite{khudo24a}. The overall dependence is not exactly parabolic because of ultrafast dynamics and quite strong quantum fluctuations  under very strong effective fields in the regions away from optimal values of $\alpha$.

\begin{figure}[htb]
    \centering
    \includegraphics[width=\columnwidth]{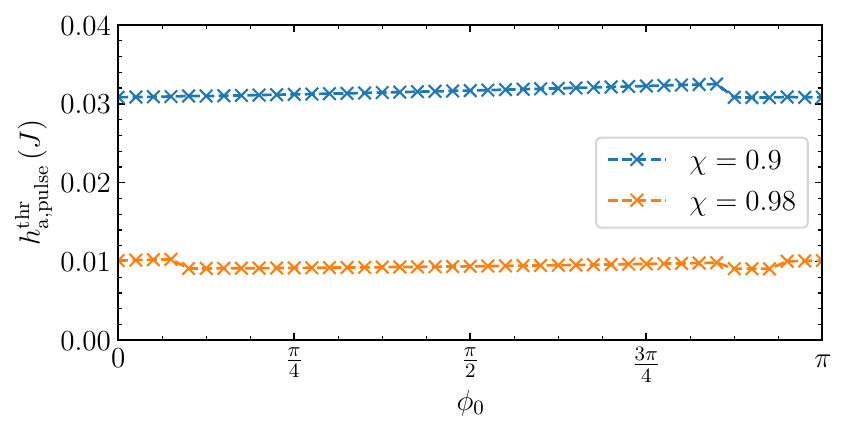}
   \caption{Threshold field dependence on initial phase with $\alpha=0.85$ and $\tau=10 \, J^{-1}$.}
    \label{fig:threshphi0}
\end{figure}

The initial phase of the pulse also crucial to catch proper spin dynamics at best resonance. For this reason we analyzed its effect on switching. Fig.\ \ref{fig:threshphi0} illustrates the threshold amplitude of the pulse dependence on initial phase. Although the dependence is very weak, the unexpected jumps also occur in this case as it was obtained recently \cite{khudo24a}. These jumps are the results of shift in switching time at some preferred oscillation under the pulse. Based on fully analyses of the pulse for other anisotropy parameters considering square lattice as well, we have chosen $\phi_0=\pi/3$ as an optimum value.

\begin{figure}[htb]
    \centering
    \includegraphics[width=\columnwidth]{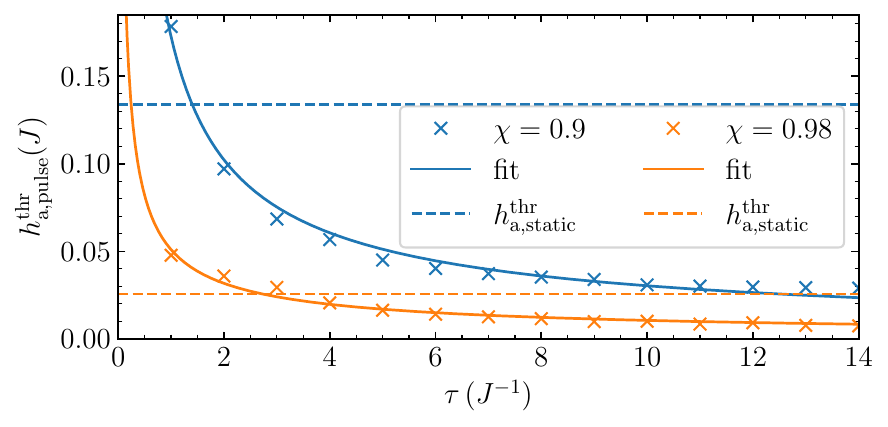}
   \caption{Threshold field dependence on pulse duration with $\alpha=0.85$ and $\phi_0=\pi/3$. Dashed horizontal lines correspond to the threshold values for alternating static magnetic field with $h\mathrm{^{thr}_{a,static}}=0.134 \,J$ and 
$h\mathrm{^{thr}_{a,static}}=0.0257 \,J$ for $\chi=0.9$ and $\chi=0.98$, respectively. The fits (solid lines) are done by power laws
 $h\mathrm{^{thr}_{a,pulse}}=a\tau^b$ with parameters $a=0.172 \, J^{b+1}$, $b=-0.755$ for $\chi=0.9$ and $a=0.05  \, J^{b+1}$, $b=-0.684$ for $\chi=0.98$. }
    \label{fig:threshtau}
\end{figure}

Finally,  the threshold field dependence on pulse duration is analyzed in Fig.\ \ref{fig:threshtau}. Obviously, longer pulses result switching at lower fields but its effect is not strong after $\tau>10 \, J^{-1}$. The fittings are done by the power low $h\mathrm{^{thr}_{a,pulse}}=a\tau^b$ and the fitting parameters $a$ and $b$ are indicated in the caption. The power low shows that the long lasting pulses can decrease the threshold values even to very small minimum as $\tau$ goes to infinity. However, we have limited the duration with $\tau=10 \, J^{-1}$ being the optimal value for our realistic pulse.


     \subsection{ Switching in square lattice by alternating fields}
  \label{app:2D}   
Here, we represent results for 2D square lattice under static and time-dependent alternating fields. General physics behind is similar to the 3D case  but with lower threshold fields as the spin gap is lower in 2D case.

\begin{figure}[htb]
    \centering
    \includegraphics[width=\columnwidth]{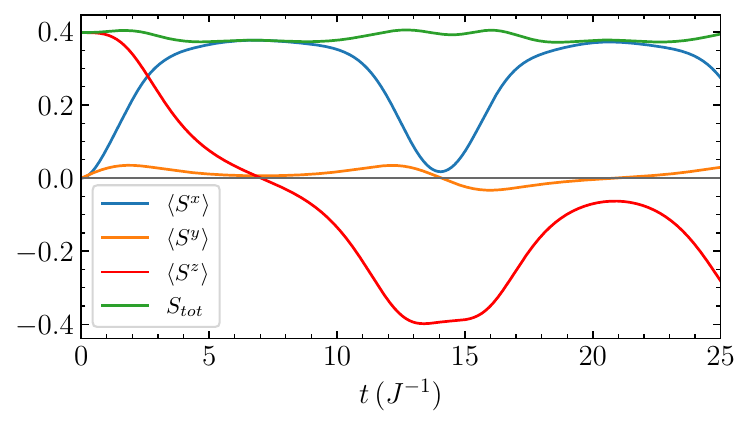}
   \caption{The dynamics of spin expectation values for $h_\mathrm{a,static}=0.08 \,J>h\mathrm{^{thr}_{a,static}}$ and $\chi=0.9$. The modules of total spin expectation value is given by $S_{tot}=\sqrt{\langle S^x\rangle^2+\langle S^y\rangle^2+\langle S^z\rangle^2}$.}
    \label{fig:sxsysz2D}
\end{figure}

Firstly, the dynamics of spin expectation values is shown in Fig.\ \ref{fig:sxsysz2D}. Overall dynamics justify the switching process in our illustration in the main part (see Fig.\ \ref{fig:ee}) with Larmor oscillations about $x$ axis and very small canting of $\langle S^y\rangle$. Clearly, the quantum oscillations also occur in the modules of total spin expectation value. 

\begin{figure}[htb]
    \centering
    \includegraphics[width=\columnwidth]{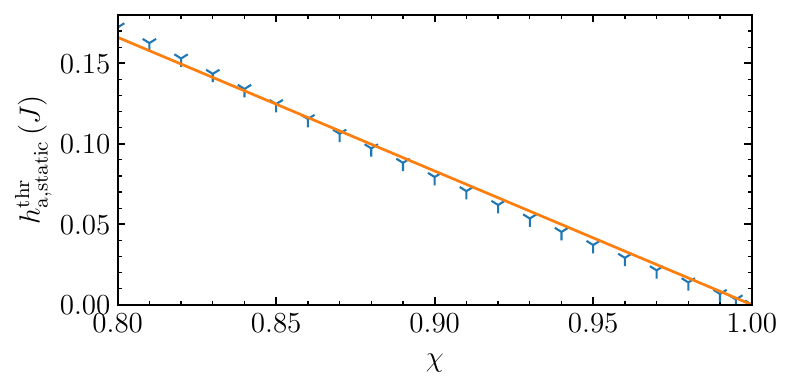}
   \caption{Threshold field dependence on anisotropy parameter. The fitting is  done by $h\mathrm{^{thr}_{a,static}}=c(1-\chi)$ where $c=0.831 \, J$.}
    \label{fig:threshchi2D}
\end{figure}
Figure \ref{fig:threshchi2D} shows threshold field  dependence on anisotropy parameter in complete analogy with the 3D case. The last minimum value correspond to the  $\chi=0.995$ with $h\mathrm{^{thr}_{a,static}}=0.0032 \, J$. This field is approximately 0.3 Tesla if one considers an appropriate antiferromagnetic exchange interaction constant. Hence, our quantum approach claims that the switching antiferromagnetic order is possible with quite low alternating fields.

\begin{figure}[htb]
    \centering
    \includegraphics[width=\columnwidth]{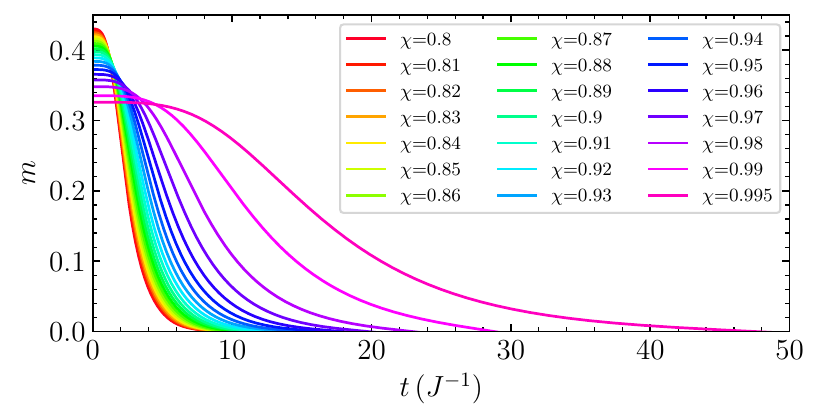}
   \caption{Dynamics of sublattice  magnetization at threshold values of alternating, static fields. }
    \label{fig:magdynamic2D}
\end{figure}
The dynamics of sablattice magnetization under threshold alternating static field is given in Fig.\ \ref{fig:magdynamic2D} for different anisotropies. One can clearly see that the weak anisotropies result slower dynamics with switching at later times, but still in THz range. 

\begin{figure}[htb]
    \centering
    \includegraphics[width=\columnwidth]{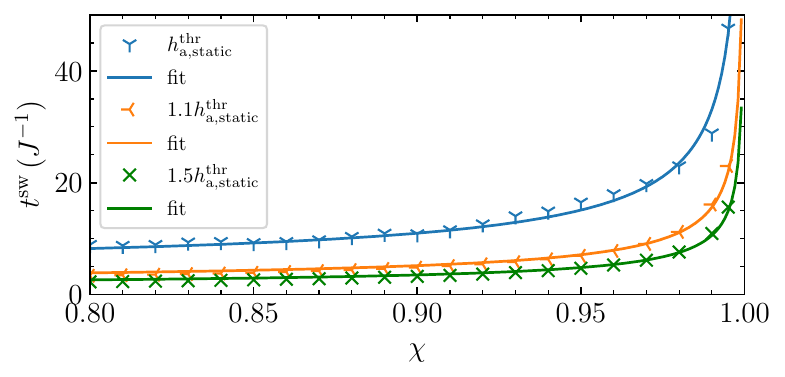}
   \caption{Switching time dependence on anisotropy. The fits are by  $t^{sw}=c/\sqrt{1-\chi}$ where $c=3.294$ for $h\mathrm{^{thr}_{a,static}}$, $c=1.553$ for $1.1 h\mathrm{^{thr}_{a,static}}$ and $c=1.054$ for $1.5 h\mathrm{^{thr}_{a,static}}$. }
    \label{fig:tswitchchi2D}
\end{figure}

The calculated switching time verses anisotropy are shown in Fig.\ \ref{fig:tswitchchi2D}. When the switching field increased by a factor of 1.1 or 1.5, we obtain earlier time switching and switching time controlled by the anisotropy of the system with inversely square root behaviour. Indeed, this is in agreement with $\sqrt{Jh_a}$ energy scale as $h_a$ has linear dependence on $\chi$.

Next, we compare the switching time of sublattice magnetization under uniform and alternating external static fields. According to the calculations in both fields with 10 percent increase from threshold value, the switching occurs slightly faster under uniform case. 
\begin{figure}[htb]
    \centering
    \includegraphics[width=\columnwidth]{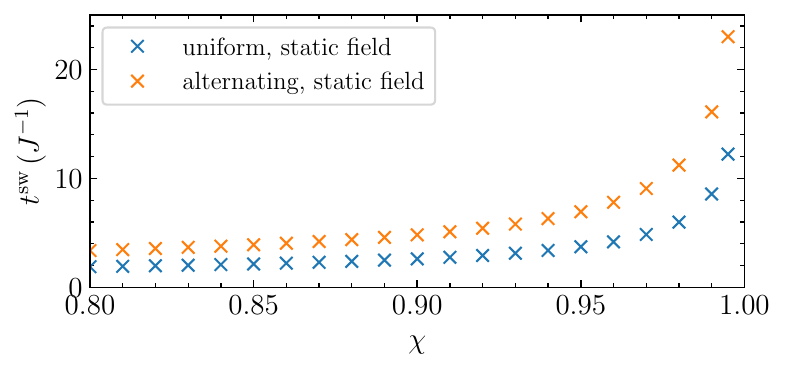}
   \caption{Switching time comparison under uniform and alternating fields.}
    \label{fig:tswitchcompare2D}
\end{figure}
The explanation of these distinction is that the strength of the threshold values for uniform fields are quite high and no additional support from effective fields are provided. However, under alternating fields, the sublattice magnetizations benefit from  effective fields and hence only initial external energy for small canting is needed. As a result, switching occurs rather late but the overall dependence on anisotropy is the same.

\begin{figure}[htb]
    \centering
    \includegraphics[width=\columnwidth]{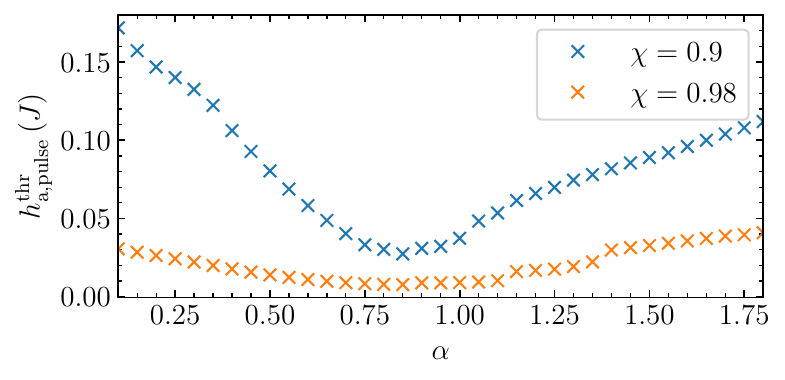}
   \caption{Switching time dependence on normalization constant. Other optimum parameters of the pulse are $\tau=10 \,J^{-1}$ and $\phi_0=\pi/3$.}
    \label{fig:treshalpha02D}
\end{figure}
The switching facilitates from time-dependent external fields at resonance with the spin gap, although the actual pulse duration is very short. To capture the full period of oscillations, we shifted the THz pulse from the time $t=0$ by $\tau=30 \, J^{-1}$ as it was mentioned in 3D case. So, here also we analyze the dynamics of sublattice magnetization in square lattice under the Gaussian pulse, given by \eqref{shortpulse}. The spin gaps in \eqref{eq:energygap} are calculated for square lattice with nearest neighbor interactions with respect to easy-axis anisotropy parameter.
Fig.\ \ref{fig:treshalpha02D} shows threshold field dependence on frequency renormalization constant. As expected, the optimal switching occurs at $\alpha\approx 0.85$ with minimum threshold field.
\begin{figure}[htb]
    \centering
    \includegraphics[width=\columnwidth]{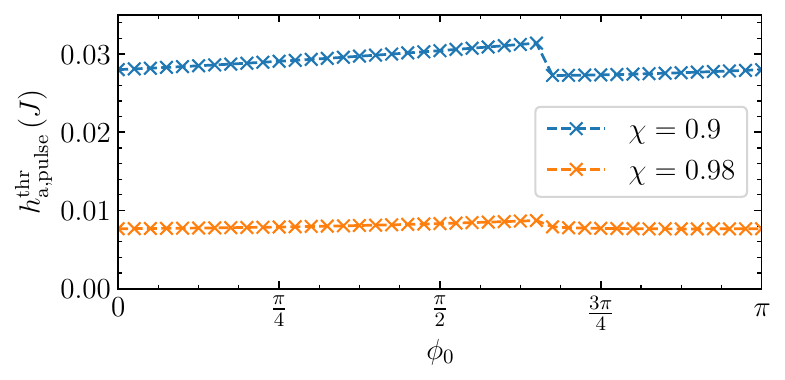}
   \caption{Switching time dependence on initial phase}
    \label{fig:treshphi02D}
\end{figure}
\begin{figure}[htb]
    \centering
    \includegraphics[width=\columnwidth]{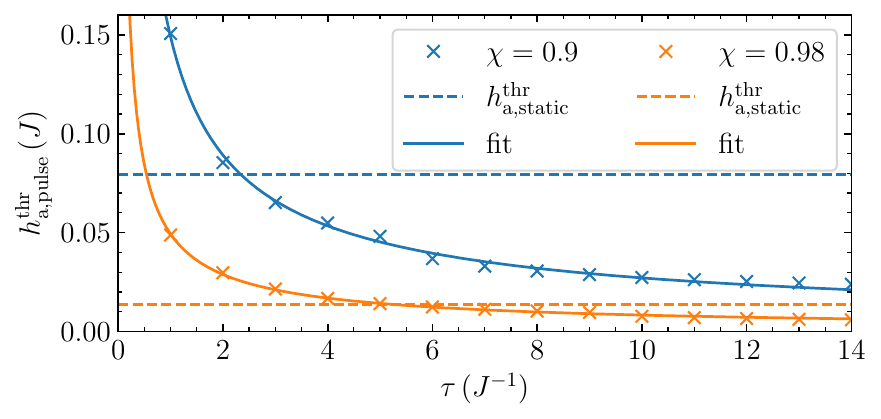}
   \caption{Switching time dependence on pulse duration. The parameters in the pulse are  $\alpha=0.85$ and $\phi_0=\pi/3$. Dashed horizontal lines correspond to the threshold values for alternating static magnetic field with $h\mathrm{^{thr}_{a,static}}=0.079 \,J$ and 
$h\mathrm{^{thr}_{a,static}}=0.0138 \,J$ for $\chi=0.9$ and $\chi=0.98$, respectively. The fits (solid lines) are done by power laws
 $h\mathrm{^{thr}_{a,pulse}}=a\tau^b$ with parameters $a=0.148 \, J^{b+1}$, $b=-0.739$ for $\chi=0.9$ and $a=0.049  \, J^{b+1}$, $b=-0.772$ for $\chi=0.98$.}
    \label{fig:treshtau02D}
\end{figure}
Lastly, Fig.\ \ref{fig:treshphi02D} and Fig.\ \ref{fig:treshtau02D} show the effect of initial phase and pulse duration on the threshold filed, respectively. The results perform a justification for the cases in 3D.

\end{document}